\documentclass[
superscriptaddress,
nofootinbib,
amsfonts,amssymb,
twocolumn,
aps,
prd,
floatfix,
longbibliography
]{revtex4-2}
		
\usepackage[colorlinks=true,urlcolor=blue,citecolor=blue,linkcolor=red]{hyperref}
\usepackage{graphicx}
\usepackage{wrapfig}
\usepackage{booktabs}
\usepackage{amsmath}
\usepackage{amssymb}
\usepackage{braket}
\usepackage[dvipsnames]{xcolor}
\usepackage[normalem]{ulem}

\usepackage{hyperref}
\usepackage{bbold}

\begin{document}
	\title{Time dilation of quantum clocks in a relativistic gravitational potential}

	\author{Tommaso Favalli}\email{tommaso.favalli@units.it}
	\affiliation{Universit\'a degli Studi di Trieste, Strada Costiera 11, I-34151 Trieste, Italy}
	\author{Augusto Smerzi}\email{augusto.smerzi@ino.it}
	\affiliation{QSTAR, INO-CNR, and LENS, Largo Enrico Fermi 2, I-50125 Firenze, Italy}

	\begin{abstract}
We study the dynamical evolution of two quantum clocks interacting with a relativistic gravitational potential. We find a time dilation effect for the clocks in agreement with the gravitational time dilation as obtained from the Schwarzschild solution in General Relativity. We perform our investigation via the Page and Wootters quantum time formalism, exploring the dynamics of clocks assuming them in both a product state and a more general (entangled) state. The gravitational redshift, as emerging from our framework, is also proposed and discussed.
	\end{abstract}
	
	\maketitle

\section{Introduction}
The general theory of relativity predicts that spacetime outside a non-rotating, spherical mass $M$ is described by the Schwarzschild metric.
In spherical coordinates, such metric is given by (c=1):
\begin{equation}\label{chiave2}
	ds^2 = \left( 1- \frac{R_S}{r} \right) dt^2 - \left( 1- \frac{R_S}{r} \right)^{-1} dr^2 + r^2 d\Omega^2
\end{equation}
where $dt$ is the coordinate time read by a far-away observer, $r$ is the Schwarzschild radial coordinate, $R_S=2GM$ is the Schwarzschild radius and $d\Omega^2$ is the metric on a unit two-sphere (see for example \cite{carrol}). 

As a consequence of (\ref{chiave2}), when considering two static clocks $A$ and $B$ at distances $r_A$ and $r_B$ from the origin of the field, we have:
\begin{equation}\label{chiave1}
\frac{\tau_B}{\tau_A} =  \left( 1- \frac{R_S}{r_B} \right)^{\frac{1}{2}} \left( 1- \frac{R_S}{r_A} \right)^{- \frac{1}{2}} 
\end{equation}
where $\tau_A$ and $\tau_B$ are the proper times measured by the clocks $A$ and $B$ respectively. Equation (\ref{chiave1}) shows that $r_B<r_A$ implies $\tau_B<\tau_A$, namely the clock $B$ results delayed with respect to clock $A$. 

In this work we show that a time dilation effect can emerge by considering two quantum clocks interacting with a \textit{relativistic gravitational potential} \cite{gravpot1,gravpot2}.  Our clocks are described by time states belonging to the complement of a bounded Hamiltonian with discrete spectrum, as introduced in \cite{pegg,chapter3,nostro}. Such clocks may have discrete or continuous time values: in the first case the complement of the Hamiltonian will be described by an Hermitian operator, while in the latter case by a POVM. We will address (together) both scenarios.

In calculating the interaction between the clocks and the relativistic gravitational potential we promote the masses of the clocks to operators using the mass-energy equivalence $m \rightarrow m + \hat{H}_{clock}$ \cite{entclockgravity,zych4,interacting,chapter7}. In this framework the coupling between the clocks and the field enters as an interaction term in the global Hamiltonian, which affects the evolution of the time states. In describing such a system we focus on the clock’s internal degrees of freedom, which are the only ones relevant to our model.

We study the dynamic of the clocks through the Page and Wootters (PaW) quantum time formalism \cite{pagewootters,wootters}. This approach to time was first proposed by D. N. Page and W. K. Wootters in 1983 and has recently gained a great deal of interest (see for example \cite{lloydmaccone,vedral,nostro2,nostro3,interacting,timedilation,foti,simile,simile2,review,review2,macconeoptempoarrivo,dirac,scalarparticles,librotommi}), including an experimental realization \cite{esp1,esp2}. In PaW theory time is a quantum degree of freedom which belongs to an ancillary Hilbert space (which we call the $C$ subspace), equipped with a suitable time observable. The dynamics of the system of interest is thus obtained with respect to the observer $C$ which we place (for convenience) at infinite distance from the source of the field. 

For a complete overview of the PaW mechanism we refer to \cite{chapter2}. We start providing a brief summary of the theory in Section II, also showing that two free clocks\footnote{Throughout the whole work we refer to \textit{free} clocks meaning clocks not perturbed by the gravitational field.} evolve synchronously with respect to the time reference $C$. 
In Section III and IV we consider instead two clocks located at different distances from the origin of the gravitational field and we show the emergence of the time dilation effect, in agreement with the Schwarzschild solution (\ref{chiave1}). We notice that this framework was already been introduced in \cite{chapter7} where the interaction of clocks with a Newtonian gravitational field was principally considered. 
Here we focus instead on the interaction with the relativistic gravitational potential, we reinterpret the results by introducing proper times for the clocks and we also study the evolution of entangled (and/or interacting) clocks. In Section V we show how the gravitational redshift can emerge in the theory. Finally, in Section VI, we give our conclusions and outlook.

\section{Evolution of free clocks}
We provide here a brief rewiev of the PaW theory following the generalization proposed in \cite{nostro,nostro2,nostro3,librotommi} and showing, as an example, the synchronous time evolution of two free clocks $A+B$ with respect to the time reference $C$. The global Hamiltonian reads:
\begin{equation}\label{h+h1}
	\hat{H}=\hat{H}_C + \hat{H}_A + \hat{H}_B
\end{equation}  
where $\hat{H}_C$, $\hat{H}_A$ and $\hat{H}_B$ are the Hamiltonians acting on $C$, $A$ and $B$, respectively. 
The key point of the PaW formalism is to consider the global quantum system in a stationary state:
\begin{equation}\label{5wdw}
	\hat{H}\ket{\Psi} = 0 .
\end{equation}
We notice and emphasize that the zero eigenvalue does not play a special role in identifying the state $\ket{\Psi}$. Indeed, up to an irrelevant global phase in the dynamics of $A+B$, the state $\ket{\Psi}$ can be identified also by imposing the constraint $\hat{H} \ket{\Psi}= \epsilon \ket{\Psi}$ with real $\epsilon$ \cite{lloydmaccone}.

\subsection{The $C$ subspace}
We assume that $\hat{H}_C$ has a discrete spectrum, with non-degenerate eigenstates having rational energy ratios.
More precisely, we consider $d_{C}$ energy states $\ket{E_i}_C$ and $E_i$ energy levels with $i=0,1,2,...,d_C -1$ such that $\frac{E_i -E_0 }{E_1 - E_0} = \frac{\beta_i}{\gamma_i}$, where $\beta_i$ and $\gamma_i$ are integers with no common factors. We obtain ($\hslash=1$):
\begin{equation}\label{5ei}
E_i = E_0 + r_i \frac{2\pi}{T_C}
\end{equation}
where $T_C =\frac{2\pi r_1}{E_1}$, $r_i = r_1\frac{\beta_i}{\gamma_i}$ for $i>1$, $r_0=0$ and $r_1$ equal to the lowest common multiple of the values of $\gamma_i$. We thus define the states 
\begin{equation}\label{5alphastateinf}
\ket{t}_C = \sum_{i=0}^{d_C-1} e^{- i E_i t}\ket{E_i}_C
\end{equation}
with $t \in \left[t_0, t_0 + T_C\right]$. These states can be used for writing the resolution of the identity in the $C$ subspace:
\begin{equation}\label{5newresolution2}
\mathbb{1}_{C} = \frac{1}{T_C} \int_{t_0}^{t_0+T_C} d t \ket{t} \bra{t} .
\end{equation}
Thanks to property (\ref{5newresolution2}) the time observable in $C$ is represented by a POVM generated by the infinitesimal operators $\frac{1}{T_C} \ket{t}\bra{t} dt$. This framework for the subspace $C$ allow us to consider any generic Hamiltonian as Hamiltonian for the $C$ subspace. Indeed, in the case of non-rational ratios of energy levels, the resolution of the identity (\ref{5newresolution2}) is no longer exact but, since any real number can be approximated with arbitrary precision by a ratio between two rational numbers, the residual terms and small corrections can be arbitrarily reduced. 

\subsection{Clocks $A$ and $B$}
We focus now on the clocks $A$ and $B$. For simplicity we take $A$ and $B$ to be equal, with equally-spaced energy spectrum. Furthermore we assume $d_A = d_B = d$ and $d_C \gg d$. The Hamiltonians are given by:
\begin{equation}
	\hat{H}_A = \hat{H}_B = \sum_{k=0}^{d-1}  \frac{2\pi}{T} k \ket{k} \bra{k}
\end{equation}
with
\begin{equation}
E^{(A)}_k = E^{(B)}_k =	\frac{2\pi}{T}k   .
\end{equation} 
In introducing the time states of the clocks, we divide the discussion assuming clocks $A$ and $B$ with discrete or with continuous time values. In the first case we have:
\begin{equation}\label{nuovo1}
	\ket{\tau_m}_A = \frac{1}{\sqrt{d}} \sum_{k=0}^{d-1} e^{-i \frac{2\pi}{T}k \tau_m} \ket{k}_A = \frac{1}{\sqrt{d}} \sum_{k=0}^{d-1} e^{-i \frac{2\pi}{d}k m} \ket{k}_A
\end{equation} 
and
\begin{equation}\label{nuovo2}
	\ket{\theta_l}_B = \frac{1}{\sqrt{d}} \sum_{k=0}^{d-1} e^{-i \frac{2\pi}{T}k \theta_l} \ket{k}_B = \frac{1}{\sqrt{d}} \sum_{k=0}^{d-1} e^{-i \frac{2\pi}{d}k l} \ket{k}_B 
\end{equation} 
where we have defined: $\tau_m = m\frac{T}{d}$ and $\theta_l = l\frac{T}{d}$ ($m,l=0,1,2,...,d-1$). 
The time values are therefore uniformly spread over the range $T$.
The states (\ref{nuovo1}) and (\ref{nuovo2}) satisfy $\braket{\tau_{m'}|\tau_m}=\delta_{m,m'}$ and $\braket{\theta_{l'}|\theta_l}=\delta_{l,l'}$. 

When considering clocks with continuous time values, we can instead introduce:
\begin{equation}\label{statitaufree3}
	\ket{\tilde{\tau}_f}_A = \sum_{k=0}^{d-1} e^{-i \frac{2\pi}{T}k \tilde{\tau}_f} \ket{k}_A = \sum_{k=0}^{d-1} e^{-i 2\pi k f} \ket{k}_A
\end{equation} 
and
\begin{equation}\label{statithetafree3}
	\ket{\tilde{\theta}_g}_B = \sum_{k=0}^{d-1} e^{-i \frac{2\pi}{T}k \tilde{\theta}_g} \ket{k}_B = \sum_{k=0}^{d-1} e^{-i 2\pi k g} \ket{k}_B 
\end{equation} 
where we have defined $\tilde{\tau}_f=fT$ and $\tilde{\theta}_g=gT$, with $f$ and $g$ taking any real values in the interval $\left[0,1\right]$. 

\subsection{Evolution of $A+B$}
Let us now look at the heart of PaW theory. Through resolution (\ref{5newresolution2}), the condensed history of the system $A+B$ can be written in the entangled global stationary state $\ket{\Psi}$, which satisfies the constraint (\ref{5wdw}). We want the clocks $A$ and $B$ to be not correlated, so we assume them in a product state, thus obtaining: 
\begin{equation}\label{statoglobaleD}
		\ket{\Psi} = \frac{1}{T_C} \int_{0}^{T_C} dt \ket{t}_C \otimes \ket{\varphi(t)}_A\otimes \ket{\phi(t)}_B
\end{equation}
where we have choosen as initial time $t_0 = 0$. In this framework the relative state (in Everett sense \cite{everett}) of $A+B$ with respect to $C$ can be obtained via conditioning:
\begin{equation}\label{defstatos}
	\ket{\varphi(t)}_A\otimes \ket{\phi(t)}_B = \braket{t|\Psi}.
\end{equation}
Note that, as mentioned before, equation (\ref{defstatos}) is the Everett \textit{relative state} definition of the subsystem $S$ with respect to the subsystem $C$. As pointed out in \cite{vedral}, this kind of projection has nothing to do with a measurement. Rather, $\ket{\varphi(t)}_A\otimes \ket{\phi(t)}_B$ is a state of $A+B$ conditioned to having the state $\ket{t}_C$ in the subspace $C$. 

For the initial state of the clocks we choose:
\begin{equation}\label{5iniziale}
\ket{\varphi(0)}_A\otimes  \ket{\phi(0)}_B \propto \sum_{k=0}^{d-1} \ket{k}_A \otimes \sum_{n=0}^{d-1} \ket{n}_B
\end{equation}
namely we consider, at time $t=0$, the clocks $A$ and $B$ to be in the time states $\ket{\tau_{m=0}}_A$ and $\ket{\theta_{l=0}}_B$ (or $\ket{\tilde{\tau}_{f=0}}_A$ and $\ket{\tilde{\theta}_{g=0}}_B$ when considering continuous time values). Thus, from equations (\ref{h+h1}), (\ref{5wdw}) and (\ref{defstatos}), it is possible demonstrate that the state of $A+B$ at generic time $t$ reads \cite{nostro}:  
\begin{multline}\label{5evoluzioneD}
		\ket{\varphi(t)}_A\otimes \ket{\phi(t)}_B = e^{-i\left(\hat{H}_A + \hat{H}_B\right)t}\ket{\varphi(0)}_A\otimes \ket{\phi(0)}_B
		\\ \propto \sum_{k=0}^{d-1} e^{-i \frac{2\pi}{T} k t} \ket{k}_A \otimes \sum_{n=0}^{d-1} e^{-i \frac{2\pi}{T} n t} \ket{n}_B 
\end{multline}
showing the Schrödinger evolution for the product state of $A+B$ with respect to the time reference $C$.

We can easily verify that $A$ and $B$ evolve synchronously. In the case of discrete time values, considering to be at time $t=m\frac{T}{d}$, we have:
\begin{multline}
\ket{\varphi(t=m\frac{T}{d})}_A\otimes \ket{\phi(t=m\frac{T}{d})}_B 
	\\ \propto \sum_{k=0}^{d-1} e^{-i \frac{2\pi}{d} k m} \ket{k}_A \otimes \sum_{n=0}^{d-1} e^{-i \frac{2\pi}{d} n m} \ket{n}_B
\end{multline}
where $A$ is clicking the state $\ket{\tau_m}_A$ and also $B$ is clicking the state $\ket{\theta_{l=m}}_B$. The same holds for clocks with continuous time values. Indeed, assuming to be at time $t=fT$, we have:
\begin{multline}\label{5evoluzioneC}
\ket{\varphi(t=fT)}_A\otimes \ket{\phi(t=fT)}_B 
	\\ \propto \sum_{k=0}^{d-1} e^{-i 2\pi k f} \ket{k}_A \otimes \sum_{n=0}^{d-1} e^{-i 2\pi n f} \ket{n}_B .
\end{multline}
Also in this case, we can see $A$ and $B$ clicking simultaneously the time states $\ket{\tilde{\tau}_f}_A$ and $\ket{\tilde{\theta}_{g=f}}_B$.


\section{Evolution of perturbed clocks}
\begin{figure}[t!] 
	\centering 
	\includegraphics [height=4cm]{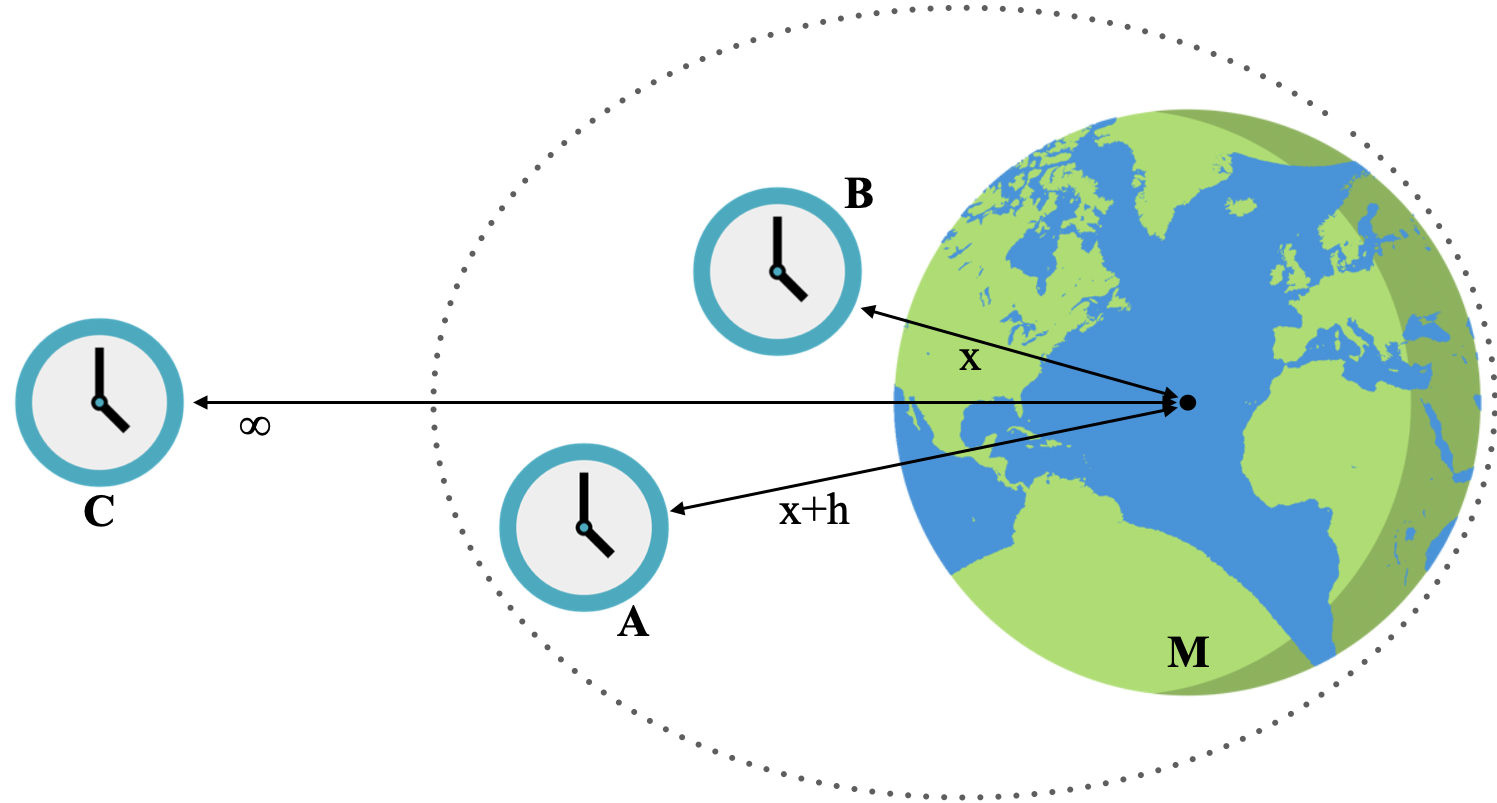} 
	\caption{The clocks $A$ and $B$ are placed in the gravitational potential at distance $x+h$ and $x$ respectively from the center of the spherical mass $M$. Their evolution is studied (via PaW formalism) with respect to the far-away observer $C$.} 
	\label{nuova1} 
\end{figure}

We consider now the case in which clocks $A$ and $B$ are placed in the gravitational field. We assume $B$ at a distance $x$ from the center of a spherical mass $M$ and $A$ placed at a distance $x+h$ (see Fig. \ref{nuova1}). When considering the relativistic gravitational potential, the energy $V$ of a clock placed at a distance $x$ from the origin of the field reads \cite{gravpot1,gravpot2}: 
\begin{equation}\label{interaction}
V = m_{clock}\left[\left(1 - \frac{2GM}{x}\right)^{\frac{1}{2}} - 1  \right] .
\end{equation}
As in \cite{entclockgravity,interacting,chapter7}, we treat the coordinate $x$ as a number and, in calculating the gravitational interaction, we promote the masses to operators using the mass–energy equivalence: $m_A \rightarrow m_A +  \hat{H}_A$ and $m_B \rightarrow m_B  +  \hat{H}_B$. Since the contributions given by the static masses would only lead to unobservable global phase factors in the evolution of the clocks, we do not consider them in the discussion. Furthermore, following \cite{entclockgravity}, we assume the clocks to follow semiclassical trajectories which are approximately static, namely with approximately zero velocity with respect to the mass $M$ and the far-away observer $C$. 

\subsection{$A$ and $B$ in the gravitational field}
The global Hamiltonian, including the interaction terms with the field, reads here:
\begin{equation}
	\begin{split}
		\hat{H} &
		= \hat{H}_C +  \hat{H}_{A}\left(1 - \frac{2GM}{x+h}\right)^{\frac{1}{2}} + \hat{H}_{B} \left(1 - \frac{2GM}{x}\right)^{\frac{1}{2}}  \\&
=	\hat{H}_C +  \hat{H'}_{A} +  \hat{H'}_{B}
	\end{split}
\end{equation}
where now
\begin{equation}\label{ha}
	\hat{H'}_{A} = \hat{H}_{A}\left(1 - \frac{2GM}{x+h}\right)^{\frac{1}{2}} = \sum_{k=0}^{d-1} \frac{2\pi}{T^{\prime\prime}} k \ket{k}\bra{k}
\end{equation}
and
\begin{equation}\label{hb}
	\hat{H'}_{B} = \hat{H}_{B}\left(1 - \frac{2GM}{x}\right)^{\frac{1}{2}} = \sum_{k=0}^{d-1} \frac{2\pi}{T'} k \ket{k}\bra{k}
\end{equation}
with $T^{\prime\prime}= T \left(1- \frac{2GM}{x+h}\right)^{ - \frac{1}{2}}$ and $T'= T \left(1- \frac{2GM}{x}\right)^{ - \frac{1}{2}}$.

We introduce again the time states. In the case of clocks with discrete time values we have:
\begin{equation}\label{nuovo3A}
	\ket{\tau_m}_A = \frac{1}{\sqrt{d}}\sum_{k=0}^{d-1} e^{-i \frac{2\pi}{T^{\prime\prime}} k \tau_m} \ket{k}_A = \frac{1}{\sqrt{d}}\sum_{k=0}^{d-1} e^{-i \frac{2\pi}{d} k m} \ket{k}_A
\end{equation}
and
\begin{equation}\label{nuovo3}
	\ket{\theta_l}_B = \frac{1}{\sqrt{d}}\sum_{k=0}^{d-1} e^{-i \frac{2\pi}{T'} k \theta_l} \ket{k}_B = \frac{1}{\sqrt{d}}\sum_{k=0}^{d-1} e^{-i \frac{2\pi}{d} k l} \ket{k}_B
\end{equation}
where we have redefined $\tau_m = m\frac{T^{\prime\prime}}{d} = m\frac{T}{d(1- \frac{2GM}{x+h})^{\frac{1}{2}}} $ and $\theta_l = l\frac{T'}{d} = l\frac{T}{d(1- \frac{2GM}{x})^{\frac{1}{2}}}$.
We notice and emphasize that the presence of the gravitational field does not change the form of the time states.
The same holds for the case of clocks with continuous time values. Indeed, in this latter case, we have:
\begin{equation}\label{taugrav}
	\ket{\tilde{\tau}_f}_A = \sum_{k=0}^{d-1} e^{-i\frac{2\pi}{T^{\prime\prime}}k \tilde{\tau}_f}\ket{k}_A = \sum_{k=0}^{d-1} e^{-i 2\pi k f}\ket{k}_A 
\end{equation}
and
\begin{equation}\label{tetagrav}
	\ket{\tilde{\theta}_g}_B = \sum_{k=0}^{d-1} e^{-i\frac{2\pi}{T'}k \tilde{\theta}_g}\ket{k}_B = \sum_{k=0}^{d-1} e^{-i 2\pi k g}\ket{k}_B 
\end{equation}
where now $\tilde{\tau}_f = fT^{\prime\prime}$ and $\tilde{\theta}_g = gT'$ with $f,g \in \left[0,1\right]$. 

We investigate the time evolution of $A+B$ in this new scenario. The global state satisfying the global constraint (\ref{5wdw}) can again be written as in (\ref{statoglobaleD}) and we assume also here the clocks starting in the product state (\ref{5iniziale}).
When the observer $C$ reads the generic time $t$, we have:
\begin{equation}\label{evrelpot}
	\begin{split}
		&\ket{\varphi(t)}_A\otimes \ket{\phi(t)}_B \propto \\&
		\sum_{k=0}^{d-1} e^{-i \frac{2\pi}{T} k t (1- \frac{2GM}{x+h})^{\frac{1}{2}}} \ket{k}_A \otimes \sum_{n=0}^{d-1} e^{-i \frac{2\pi}{T} n t(1- \frac{2GM}{x})^{\frac{1}{2}}} \ket{n}_B.
	\end{split}
\end{equation}

In the case of clocks with discrete time values, considering to be at time $t=m \frac{T}{d}$, equation (\ref{evrelpot}) becomes:
\begin{equation}
	\begin{split}
		 &\ket{\varphi(t=m \frac{T}{d})}_A\otimes \ket{\phi(t=m \frac{T}{d})}_B  \propto \\& \sum_{k=0}^{d-1} e^{-i \frac{2\pi}{d} k m(1- \frac{2GM}{x+h})^{\frac{1}{2}}} \ket{k}_A \otimes \sum_{n=0}^{d-1} e^{-i \frac{2\pi}{d} n m(1- \frac{2GM}{x})^{\frac{1}{2}}} \ket{n}_B .
		\end{split}
\end{equation}
This implies that, when $A$ is clicking the time state $\ket{\tau_{m^{\prime\prime}}}$, clock $B$ has clicked a number of states:
\begin{equation}\label{1}
	m' 	 =   m^{\prime\prime} \left( 1- \frac{2GM}{x} \right)^{\frac{1}{2}} \left( 1- \frac{2GM}{x+h} \right)^{- \frac{1}{2}} 
\end{equation}
which is in agreement with the time dilation between two clocks at a (radial) distance $h$ from each other as obtained from the Schwarzschild solution (see equation (\ref{chiave1})). 

Similarly, when considering clocks with continuous time values, assuming to be at $t=fT$, equation (\ref{evrelpot}) becomes: 
\begin{equation}\label{5evoluzioneC2.2}
	\begin{split}
	  &\ket{\varphi(t=fT)}_A\otimes \ket{\phi(t=fT)}_B \propto \\&  \sum_{k=0}^{d-1} e^{-i 2\pi k f(1- \frac{2GM}{(x+h)c^2})^{\frac{1}{2}}} \ket{k}_A \otimes \sum_{n=0}^{d-1} e^{-i 2\pi n f(1- \frac{2GM}{xc^2})^{\frac{1}{2}}} \ket{n}_B .
	\end{split}
\end{equation}
This implies that, when $A$ clicks the time state $\ket{\tilde{\tau}_{f^{\prime\prime}}}_A$, the clock $B$ is clicking the time state $\ket{\tilde{\theta}_{g=f'}}_B$ with:
\begin{equation}\label{2}
	f' 	 =   f^{\prime\prime} \left( 1- \frac{2GM}{x} \right)^{\frac{1}{2}} \left( 1- \frac{2GM}{x+h} \right)^{- \frac{1}{2}} .
\end{equation}

We can also put clock $A$ at infinite distance from the mass. By taking $h \longrightarrow \infty$, equations (\ref{1}) and (\ref{2}) become: $m'=m^{\prime\prime}\left( 1- \frac{2GM}{x} \right)^{\frac{1}{2}}$ and $f'=f^{\prime\prime}\left( 1- \frac{2GM}{x} \right)^{\frac{1}{2}}$,
again in agreement with the temporal term in the Schwarzschild metric
\begin{equation}
	d\tau = \left(1 - \frac{2GM}{x}\right)^{\frac{1}{2}} dt
\end{equation}
where $d\tau$ is the infinitesimal proper time read within the field and $dt$ is the time read by a far-away observer.

\subsection{Introducing the proper time}
In this paragraph we briefly reinterpret what we have just discussed by introducing the proper time read by the clocks $A$ and $B$. In equation (\ref{evrelpot}) we found that, when the observer $C$ reads the generic time $t$, the states of $A+B$ evolve according to:
\begin{equation}\label{pt1}
	\begin{split}
		&\ket{\varphi(t)}_A\otimes \ket{\phi(t)}_B \propto \\&
		\sum_{k=0}^{d-1} e^{-i \frac{2\pi}{T} k t (1- \frac{2GM}{x+h})^{\frac{1}{2}}} \ket{k}_A \otimes \sum_{n=0}^{d-1} e^{-i \frac{2\pi}{T} n t(1- \frac{2GM}{x})^{\frac{1}{2}}} \ket{n}_B.
	\end{split}
\end{equation}
We can now define the proper times read by clocks $A$ and $B$ as: $\tau_A(t)=t (1- \frac{2GM}{x+h})^{\frac{1}{2}}$ and $\tau_B(t)=t (1- \frac{2GM}{x})^{\frac{1}{2}}$. Equation (\ref{pt1}) can thus be rewritten:
\begin{multline}\label{pt2}
		\ket{\varphi(t)}_A\otimes \ket{\phi(t)}_B \propto \\
		\sum_{k=0}^{d-1} e^{-i \frac{2\pi}{T} k \tau_A(t)} \ket{k}_A \otimes \sum_{n=0}^{d-1} e^{-i \frac{2\pi}{T} n \tau_B(t)} \ket{n}_B
\end{multline}
where it is manifest that, in the product state of $A+B$, each clock evolves according to its proper time. 

This result can also be summarized by writing:
\begin{multline}
		\ket{\varphi(t)}_A\otimes \ket{\phi(t)}_B = e^{-i\ \left(\hat{H'}_A +  \hat{H'}_B\right)t }\ket{\varphi(0)}_A \otimes \ket{\phi(0)}_B \\= e^{-i\hat{H}_A \tau_A(t) }\ket{\varphi(0)}_A \otimes e^{-i\hat{H}_B \tau_B(t) }\ket{\phi(0)}_B
\end{multline}
clearly showing that the effect of interaction with gravitational potential can be interpreted as (properly) dilating time, while leaving the clocks energy unchanged.

\section{Evolution of entangled clocks}
In this Section we study the evolution of clocks $A$ and $B$ in a generic (i.e. not product) state. We consider the clocks in the gravitational potential, with $B$ at a distance $x$ from the center of the mass $M$ and $A$ at a distance $x+h$, as in the previous Section.
The global state satisfying the global constraint (\ref{5wdw}) can be written here: 
\begin{equation}\label{statoglobaleD2}
	\begin{split}
		\ket{\Psi} &= \frac{1}{T_C} \int_{0}^{T_C} dt \ket{t}_C \otimes \ket{\psi(t)}_{AB} 
		\\&= \frac{1}{T_C} \int_{0}^{T_C} dt \ket{t}_C \otimes e^{-i\hat{H}_{AB}t} \ket{\psi(0)}_{AB}
	\end{split}
\end{equation}
where $\hat{H}_{AB}$ and $\ket{\psi(0)}_{AB}$ are the Hamiltonian and the initial state referring to the subsystem $A+B$ of the clocks. We start by exploring the time evolution of the clocks in this new case and then we also consider the case in which an interaction term between the clocks is present.

\subsection{Clocks $A$ and $B$ in a generic state}
We take the initial state of $A+B$ appearing in (\ref{statoglobaleD2}) as the generic state in the energy eigenbasis:
\begin{equation}
	\ket{\psi(0)}_{AB} = \sum_{n=0}^{d-1} \sum_{k=0}^{d-1} c_{nk} \ket{n}_A \otimes \ket{k}_B
\end{equation}
with $\sum_{n,k}|c_{nk}|^2=1$, and we calculate its evolution through the Hamiltonian
\begin{equation}
		\hat{H}_{AB} =  \hat{H}_{A}\left(1 - \frac{2GM}{x+h}\right)^{\frac{1}{2}} + \hat{H}_{B} \left(1 - \frac{2GM}{x}\right)^{\frac{1}{2}} .
\end{equation}
Using the results of paragraph III.B, we obtain:
\begin{equation}\label{evolutionentangled}
	\begin{split}
		\ket{\psi(t)}_{AB} &= e^{-i\hat{H}_{AB}t} \ket{\psi(0)}_{AB}
		\\& = e^{-i \left(  \hat{H}_{A}\left(1 - \frac{2GM}{x+h}\right)^{\frac{1}{2}} + \hat{H}_{B} \left(1 - \frac{2GM}{x}\right)^{\frac{1}{2}} \right)t}  \ket{\psi(0)}_{AB}
		\\& = e^{-i \hat{H}_{A}\tau_A(t) } e^{-i \hat{H}_{B}\tau_B(t)} \ket{\psi(0)}_{AB}
		\\& = \sum_{n=0}^{d-1} \sum_{k=0}^{d-1} c_{nk} e^{-i \frac{2\pi}{T}n \tau_A(t) } \ket{n}_A \otimes  e^{-i \frac{2\pi}{T}k \tau_B(t)} \ket{k}_B
		\\& =  \sum_{n=0}^{d-1} \sum_{k=0}^{d-1} c_{nk} e^{-i \frac{2\pi}{T} \left(n\tau_A(t) + k\tau_B(t) \right)} \ket{n}_A \otimes \ket{k}_B
	\end{split}
\end{equation}
where $\tau_A(t)$ and $\tau_B(t)$ are the proper times read by clocks $A$ and $B$ respectively when the observer in $C$ read the time $t$. From equation (\ref{evolutionentangled}) we can easily see how, within a generic state of $A+B$, each term in the superposition acquires a phase proportional to the sum of integer multiples of proper times read by the two clocks. 

To better understand this kind of evolution we can look at a simple example by taking d=2. Namely we consider, for $A$ and $B$, the simplest choice of clock: a qubit. In Sections II and III we assumed the ground state of the Hamiltonian with zero energy, but the framework can be easily generalized for energy translations. We therefeore take:
\begin{equation}
	\hat{H}_A = \hat{H}_B = \frac{\omega}{2} \hat{\sigma}_z 
\end{equation}
leading to the Hamiltonian:
\begin{equation}
	\hat{H}_{AB} = \frac{\omega}{2} \hat{\sigma}^{(A)}_z \left(1 - \frac{2GM}{x+h}\right)^{\frac{1}{2}} +  \frac{\omega}{2} \hat{\sigma}^{(B)}_z \left(1 - \frac{2GM}{x}\right)^{\frac{1}{2}} .
\end{equation}
The (initial) generic state of two qubit can be written:
\begin{equation}\label{qubitiniziale}
		\ket{\psi(0)}_{AB} = \alpha \ket{00} + \beta \ket{01} + \gamma \ket{10} + \delta \ket{11}
\end{equation}
with $|\alpha|^2 + |\beta|^2+|\gamma|^2+|\delta|^2=1$. In the state (\ref{qubitiniziale}) the first position in the kets refers to clock $A$ and the second to clock $B$. We can now write:
\begin{equation}
	\begin{split}
		e^{-i \frac{\omega}{2}  \hat{\sigma}^{(A)}_z \left(1 - \frac{2GM}{x+h}\right)^{\frac{1}{2}} t} 
		= \left(\begin{matrix}
			e^{-i\frac{\omega}{2}\tau_A(t)}&0\\0&e^{i\frac{\omega}{2}\tau_A(t)}
		\end{matrix}\right)
	\end{split}
\end{equation}
and similarly
\begin{equation}
	\begin{split}
		e^{-i \frac{\omega}{2}  \hat{\sigma}^{(B)}_z \left(1 - \frac{2GM}{x+h}\right)^{\frac{1}{2}} t} 
		= \left(\begin{matrix}
			e^{-i\frac{\omega}{2}\tau_B(t)}&0\\0&e^{i\frac{\omega}{2}\tau_B(t)}
		\end{matrix}\right) .
	\end{split}
\end{equation}
Thus, the time evolution of (\ref{qubitiniziale}) can be easily calculated:
\begin{multline}\label{even1}
		\ket{\psi(t)}_{AB} = \alpha e^{- i \frac{\omega}{2}\left( \tau_A(t) + \tau_B(t)\right)}\ket{00} +\\+ \beta e^{- i \frac{\omega}{2}\left( \tau_A(t) - \tau_B(t)\right)}\ket{01} + \\ 
		+  \gamma e^{ i \frac{\omega}{2}\left( \tau_A(t) - \tau_B(t)\right)}\ket{10} +\\+ \delta e^{ i \frac{\omega}{2}\left( \tau_A(t) + \tau_B(t)\right)}\ket{11}
\end{multline}
where it is manifest that the phases acquired by the various states in the superposition depend on sums or differences (remember that here the energy spectrum of $A$ and $B$ has values $\pm \frac{\omega}{2}$) of the two clocks' proper times.

Finally, we rewrite (\ref{even1}) as 
\begin{equation}\label{even2}
	\ket{\psi(t)}_{AB} = \alpha(t)\ket{00} + \beta(t)\ket{01} 
+  \gamma(t)\ket{10} + \delta(t)\ket{11} 
\end{equation}
and we calculate the concurrence $C(\psi(t))=2|\alpha(t) \delta(t) - \beta(t) \gamma(t)|$, to keep track of the measure of entanglement over time \cite{woottersconc,conc2}. We find:
\begin{equation}
	\begin{split}
		C(\psi(t)) &= 2\left| \alpha(t) \delta(t) - \beta(t) \gamma(t)\right|
		\\&= 2| \alpha e^{- i \frac{\omega}{2}\left( \tau_A(t) + \tau_B(t)\right)} \delta  e^{ i \frac{\omega}{2}\left( \tau_A(t) + \tau_B(t)\right)} \\& -   \beta e^{- i \frac{\omega}{2}\left( \tau_A(t) - \tau_B(t)\right)}    \gamma e^{ i \frac{\omega}{2}\left( \tau_A(t) - \tau_B(t)\right)} |
		\\& =2\left| \alpha \delta - \beta\gamma\right| = C(\psi(0))
	\end{split}
\end{equation} 
showing (as expected) that the interaction with the gravitational potential is not able to change the measure of the entanglement present in the initial state.

\subsection{Entanglement of interacting clocks}
In order to observe a change in the measure of entanglement during the clocks dynamics, it is necessary to introduce a term of interaction between them. For this reason, in this paragraph, we study the time evolution of $A$ and $B$ described by two qubit with the addition of a simple interaction term $\hat{H}_{int} = \epsilon \hat{H}_A \otimes \hat{H}_B$ in the Hamiltonian $\hat{H}_{AB}$. This form of interaction is obtained for example if one considers the clocks interacting through Newtonian gravity, by taking $\epsilon = -G/h$ with $h$ the distance between $A$ and $B$ (see \cite{entclockgravity,interacting,librotommi}).

The initial state of clocks $A$ and $B$ is again given by (\ref{qubitiniziale}) and we calculate its evolution through the Hamiltonian:
\begin{multline}
	\hat{H}_{AB} =  \frac{\omega}{2} \hat{\sigma}^{(A)}_z \left(1 - \frac{2GM}{x+h}\right)^{\frac{1}{2}} +  \frac{\omega}{2} \hat{\sigma}^{(B)}_z \left(1 - \frac{2GM}{x}\right)^{\frac{1}{2}} + 
	\\ + \frac{\epsilon \omega^2}{2}  \hat{\sigma}^{(A)}_z \otimes  \hat{\sigma}^{(B)}_z .
\end{multline}
Thus, the state of $A+B$ at generic time $t$ reads:
\begin{widetext}
\begin{equation}
	\ket{\psi(t)}_{AB} = e^{-i \frac{\omega}{2}\tau_A(t) \hat{\sigma}^{(A)}_z}e^{-i \frac{\omega}{2}\tau_B(t) \hat{\sigma}^{(B)}_z}e^{-i  \frac{\epsilon \omega^2}{2} t  \hat{\sigma}^{(A)}_z \otimes  \hat{\sigma}^{(B)}_z  } \ket{\psi(0)}_{AB}
\end{equation}
\end{widetext}
which leads to the result
\begin{multline}\label{even3}
		\ket{\psi(t)}_{AB} = \alpha e^{- i \frac{\omega}{2}\left( \tau_A(t) + \tau_B(t)\right)} e^{-i  \frac{\epsilon \omega^2}{2} t}\ket{00} +\\+ \beta e^{- i \frac{\omega}{2}\left( \tau_A(t) - \tau_B(t)\right)}e^{i  \frac{\epsilon \omega^2}{2} t}\ket{01} + \\ 
		+  \gamma e^{ i \frac{\omega}{2}\left( \tau_A(t) - \tau_B(t)\right)}e^{i  \frac{\epsilon \omega^2}{2} t}\ket{10} +\\+ \delta e^{ i \frac{\omega}{2}\left( \tau_A(t) + \tau_B(t)\right)}e^{-i  \frac{\epsilon \omega^2}{2} t}\ket{11} 
\end{multline}
where we can see that, in addition to the phases containing the proper times of the clocks, the time $t$ read by the observer $C$ also explicitly appears. 

We can now again rewrite equation (\ref{even3}) as:
\begin{equation}\label{even4}
	\ket{\psi(t)}_{AB} = \alpha(t)\ket{00} + \beta(t)\ket{01} 
	+  \gamma(t)\ket{10} + \delta(t)\ket{11} 
\end{equation}
and calculate the concurrence $C(\psi(t))=2|\alpha(t) \delta(t) - \beta(t) \gamma(t)|$. We obtain:
\begin{equation}\label{concfinale}
	C(\psi(t))=2\left| \alpha \delta e^{-i\epsilon \omega^2 t} - \beta \gamma e^{i\epsilon \omega^2 t}\right| 
\end{equation}
where we can immediately observe that the dependence on proper times vanishes, while the contributions associated with the time $t$ persist. Equation (\ref{concfinale}) thus shows that the concurrence is in general oscillating with time $t$. If for example we choose $\alpha\delta = \beta\gamma=\frac{1}{4}$, we obtain $C(\psi(t)) = |\sin{(\epsilon \omega^2 t)}|$, displaying an initially unentangled state, whose concurrence subsequently oscillates between zero and nonzero values as time $t$ evolves.

Finally, we observe that, given the form of our interaction, it is sufficient for one of the coefficients $\alpha, \beta, \gamma, \delta$ to be equal to zero in order for the concurrence to remain constant in time. An example is provided by the Bell states, for which we have $\alpha\delta = \pm \frac{1}{2}, \beta\gamma = 0$ or $\alpha\delta = 0, \beta\gamma = \pm \frac{1}{2}$. In such cases, it is straightforward to show that the entanglement remains maximal at all times, with $C(\psi(t)) = 1$.

\section{Gravitational redshift}
We derive here the gravitational redshift as emerging in our framework. For this Section we introduce~$c \ne 1$. 

We consider $A$ and $B$ both placed in the gravitational potential at distance $x+h$ and $x$ respectively from the origin of the field (the Hamiltonians are given by (\ref{ha}) and (\ref{hb})) and we assume an observer in $A$ receiving a light signal emitted in $B$.
We assume the frequecy of the light signal as proportional to the spacing between two neighboring energy levels of the clocks, namely $1/T$ for a free clock. 
The observer $A$ can thus read the frequency $\nu_O$ coming from $B$ and compare it with his own spectrum, that is: $\delta \nu = \nu_O - \nu = \frac{1}{T^{'}} -  \frac{1}{ T^{\prime\prime}} $,
leading to
\begin{equation}
\delta \nu = \frac{1}{T} \left[    \left(1- \frac{2GM}{xc^2}\right)^{\frac{1}{2}}  -   \left(   1- \frac{2GM}{(x+h)c^2}\right)^{ \frac{1}{2}}      \right].
\end{equation}
At the first order of approximation, when $\frac{2GM}{xc^2} \ll 1$, we therefore have:
\begin{equation}
	\delta \nu  \simeq \frac{1}{T} \left[ \left(1 - \frac{GM}{xc^2}\right) -  \left(1 - \frac{GM}{(x+h)c^2}\right)\right] 
\end{equation}
that, for $h \ll x$, becomes
\begin{equation}
	\delta \nu \simeq \frac{1}{T}  \frac{GM}{c^2} \left( \frac{1}{x+h} - \frac{1}{x}  \right) \simeq - \frac{1}{T} \frac{GMh}{x^2c^2}  . 
\end{equation}
Writing now the gravitational acceleration $a=\frac{GM}{x^2}$ and neglecting terms of the order $\sim \left(\frac{GM}{xc^2}\right)^2$, we obtain
\begin{equation}\label{frequenzaD}
	\frac{\delta \nu}{\nu} \simeq - \frac{ah}{c^2} .
\end{equation}
Equation (\ref{frequenzaD}) is in agreement with what is measured in experiments on Earth (see for example \cite{misuragravita}). It clearly holds when considering the spacing between any two energy levels and not only between two neighbors.



\section{Conclusions}
In this work, through the PaW theory, we examined the time evolution of two quantum clocks ($A$ and $B$) when interacting with a relativistic gravitational potential. We have performed our investigation in the case clocks with discrete and continuous time values. In both cases we first verified that, in the absence of the field, the two clocks evolved synchronously. Then, promoting the mass of the clocks to operator, we introduced the interaction with the field and we found a time dilation effect for the time states of the clocks in agreement with the Schwarzschild solution (\ref{chiave2}). By introducing the proper time for the clocks, we thus showed that the effect of interaction with the gravitational potential can be interpreted as (properly) dilating time, while leaving the clocks energy unchanged. The evolution of entangled clocks was also studied, revealing that the entanglement measure remains unaffected by the gravitational field alone, and changes only in the presence of an interaction term between the clocks.
The expression for the gravitational redshift was finally derived and discussed.

As subject of a future work, we propose to introduce also the spatial degree of freedom in the discussion. Through the interaction of a quantum ruler with the relativistic gravitational potential, we will be able (hopefully) to show that the possible outomes of a position measurement, made on the ruler placed in the gravitational field, are modified in agreement with the gravitational lengths stretching as obtained from the Schwarzschild metric. Through the study of a time-evolving ruler, we thus would derive and discuss the probability that the ruler connects events in spacetime. We notice that this proposal would be carried out in a fully relational approach, moving the discussion into the more general context of quantum reference frames \cite{librotommi,nostro3}. 

In conclusion, we emphasize that the choice of treating the distance $x$ from the origin of the field as a number is only an approximation, useful to show the power of the framework: it can not be the ultimate solution (especially in future developments of the theory where a quantum ruler should be introduced). 
We thus propose to move away from this approximation in the future by providing a framework where the distance of the clocks from the origin of the field is treated as an operator.


\section*{Acknowledgements}
The authors thank the European Commission through the H2020 QuantERA ERA-NET Cofund
in Quantum Technologies project \lq\lq MENTA\rq\rq. T.F. thanks the Project \lq\lq National Quantum Science and Technology Institute – NQSTI\rq\rq Spoke 3 \lq\lq Atomic, molecular platform for quantum technologies\rq\rq. T.F. also thanks Lapo Casetti, Andrea Trombettoni and Lorenzo Bartolini for useful discussions. A.S. thanks Vasilis Kiosses for discussions.



\end{document}